\documentclass[conference,9pt]{IEEEtran}
\IEEEoverridecommandlockouts
\usepackage{multirow}
\usepackage{cite}
\usepackage{amsmath,amssymb,amsfonts}
\usepackage{algorithmic}
\usepackage{graphicx}
\usepackage{textcomp}
\usepackage{xcolor}
\usepackage{booktabs}
\usepackage{makecell}
\def\BibTeX{{\rm B\kern-.05em{\sc i\kern-.025em b}\kern-.08em
    T\kern-.1667em\lower.7ex\hbox{E}\kern-.125emX}}

\usepackage{newtxtext}
\usepackage{newtxmath}
\usepackage[T1]{fontenc}
\usepackage{mathptmx}
\usepackage[a-1b]{pdfx}

\begin{document}

\title{DeSTA2: Developing Instruction-Following Speech Language Model Without Speech Instruction-Tuning Data
}

\author{ \IEEEauthorblockN{Ke-Han Lu$^1$, Zhehuai Chen$^2$, Szu-Wei Fu$^2$, Chao-Han Huck Yang$^2$,\\Jagadeesh Balam$^2$, Boris Ginsburg$^2$, Yu-Chiang Frank Wang$^2$, Hung-yi Lee$^1$}
\IEEEauthorblockA{$^1$Graduate Institute of Communication Engineering, National Taiwan University 
  $^2$NVIDIA}
}

\maketitle

\begin{abstract}

Recent end-to-end speech language models (SLMs) have expanded upon the capabilities of large language models (LLMs) by incorporating pre-trained speech models. However, these SLMs often undergo extensive speech instruction-tuning to bridge the gap between speech and text modalities. This requires significant annotation efforts and risks catastrophic forgetting of the original language capabilities. In this work, we present a simple yet effective automatic process for creating speech-text pair data that carefully injects speech paralinguistic understanding abilities into SLMs while preserving the inherent language capabilities of the text-based LLM. Our model demonstrates general capabilities for speech-related tasks without the need for speech instruction-tuning data, achieving impressive performance on Dynamic-SUPERB and AIR-Bench-Chat benchmarks. Furthermore, our model exhibits the ability to follow complex instructions derived from LLMs, such as specific output formatting and chain-of-thought reasoning. Our approach not only enhances the versatility and effectiveness of SLMs but also reduces reliance on extensive annotated datasets, paving the way for more efficient and capable speech understanding systems. \footnote{https://kehanlu.github.io/DeSTA2}




\end{abstract}


{\begin{figure*}[ht]
\centerline{\includegraphics[width=0.87\linewidth]{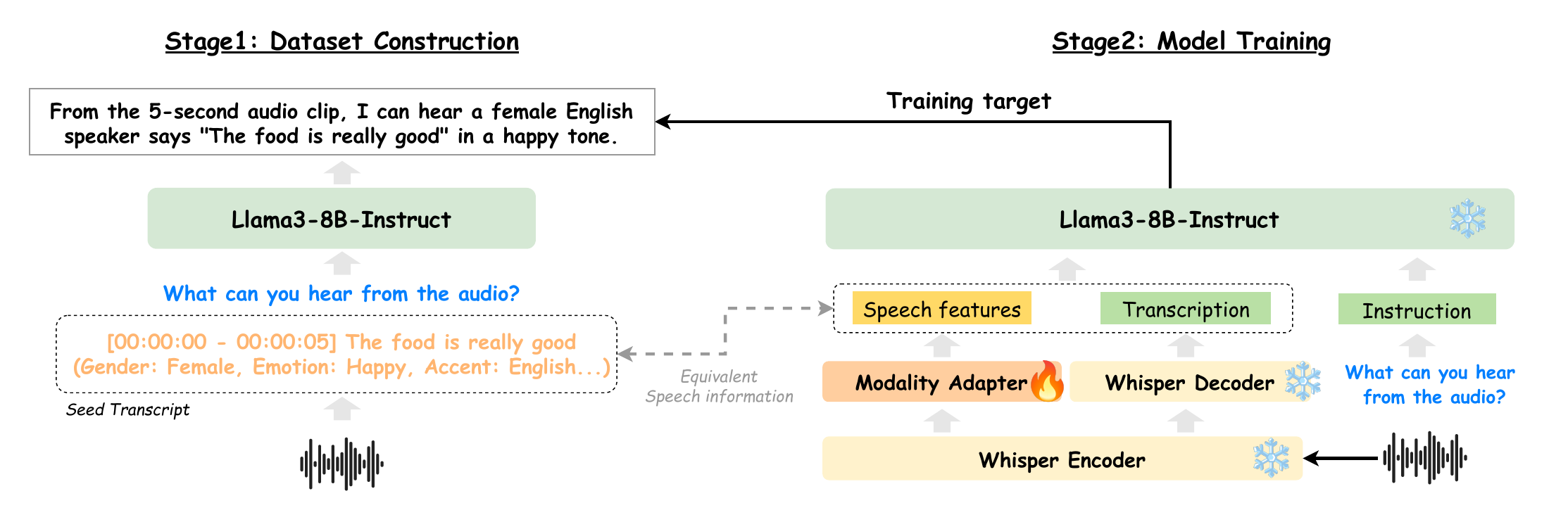}}
\vspace{-1em}
\caption{\textit{(Left)} \textbf{Dataset construction} We feed seed transcript and prompt to generate response as training target. \textit{(Right)} \textbf{Model training} The end-to-end model learns to generate same response based on speech features and text transcription.}
\vspace{-1em}
\label{fig:model}
\end{figure*}}

\section{Introduction}


Large language models (LLMs) \cite{achiam2023gpt,dubey2024llama,anil2023palm,yang2024qwen2} have demonstrated remarkable capabilities in natural language processing and comprehension, leveraging vast amounts of training data to achieve general proficiency. Building on this success, researchers have begun exploring the extension of these capabilities to speech processing tasks. This emerging field focuses on developing speech language models (SLMs) \cite{gong2024listen,gong_ltuas,tang2023salmonn,chu2023qwen,chu2024qwen2,wang2023blsp,hu2024wavllm,huang2023dynamic,kuan2024speech,huang2024audiogpt,huang2024speechcaps,das2024speechverse} that can interpret textual instructions and process speech input, enabling them to perform a wide range of speech-related tasks. The challenge can be decomposed into two fundamental abilities: speech understanding and instruction-following. Typically, SLMs integrate pre-trained speech models with large language models, aiming to leverage the strengths of both components. 

Developing SLMs presents significant challenges, particularly in constructing suitable speech-text training pairs to bridge the gap between text and speech modalities.
While training strategies vary across studies, most require a substantial speech instruction-tuning stage to enable broad understanding of speech tasks. One common approach is utilizing textual metadata or labels from existing speech datasets, often employing human annotators to create task-specific instructions \cite{tang2023salmonn,huang2023dynamic} or using text-based LLMs to generate open-ended question-answer pairs \cite{gong_ltuas,gong2024listen,chu2023qwen,chu2024qwen2,hu2024wavllm,das2024speechverse}. However, these approaches have severe limitations. The quality of the resulting dataset heavily depends on the annotators or system prompts, inevitably introducing annotation bias or ambiguity into the training data, such as specific output formats or word patterns. Consequently, the model must not only learn to understand speech features but also contend with annotation biases. We believe this could potentially lead to task overfitting \cite{tang2023salmonn} in SLMs, where the model becomes overly specialized in producing specific forms of output or performing particular tasks.
Furthermore, given that text-based LLMs already possess the ability to understand and follow diverse instructions, it's worth questioning whether a massive speech instruction-tuning stage is truly necessary for developing SLMs. We should consider if it's possible to enhance the model's speech comprehension capabilities without compromising the original versatility of the LLM. In this work, we demonstrate that the answer is "\textbf{yes}". Our approach addresses two critical challenges simultaneously: mitigating the forgetting problem often associated with specialized finetuning and reducing the need for extensive human effort in data preparation.

In this work, we introduce a dataset construction approach that combines two innovations:
First, we leverage the text-based LLM in our SLM to generate speech-text pairs using textual metadata collected from speech. This crucial design minimizes the textual mismatch between the original LLM and new speech-text training data, allowing the model to focus primarily on learning speech information. These speech metadata covers a wide range of speech attributes, which are either from human annotation or extracted by several specialized speech models.
Second, motivated by the idea of learning general speech concepts from speech captions \cite{lu24c_interspeech}, we employ a single prompt—\textit{"What can you hear from the audio?"} for both data construction and model training, which significantly reduces the need for task-specific annotation.
We call this carefully designed \textbf{de}scriptive \textbf{s}peech-\textbf{t}ext \textbf{a}lignment approach DeSTA2.

DeSTA2 achieves superior or comparable results on Dynamic-SUPERB\cite{huang2023dynamic} and AIR-Bench-Chat\cite{yang2024air} \textit{without} training on any task-specific instruction pairs, surpassing previous models that relied heavily on instruction-tuning dataset. This success highlights the effectiveness of our approach in aligning speech and text modalities while avoiding the forgetting problem.




We summarize our contribution as follows.
\begin{enumerate}
    \item 
    
    We challenge the prevailing belief that a massive speech instruction-tuning stage is required for developing instruction-following SLMs. Instead, we demonstrate that a speech caption dataset with a proper data construction process and rich speech metadata can already build a robust and versatile SLM from a powerful text-based LLM.

    \item DeSTA2 achieves impressive performance on both Dynamic-SUPERB and AIR-Bench-Chat, which require strong instruction-following and speech understanding abilities. Our approach surpasses previous models that relied heavily on instruction-tuning and complicated training strategies.
    \item Our model retains the advanced reasoning capabilities of the original text-based LLM, including the ability to follow complex instructions and engage in chain-of-thought reasoning. This versatility extends well beyond the scope of the speech-text training data, highlighting a significant advantage of our method over previous approaches. We will release our model in the future.
    

    

\end{enumerate}

\section{Related work}


Recent research on SLMs has focused on bridging modality gaps by creating speech-text pairs. Many approaches employ multi-stage curriculum learning strategies to gradually develop speech processing abilities across various tasks \cite{tang2023salmonn,chu2023qwen,chu2024qwen2,hu2024wavllm,lu24c_interspeech}. For instance, Qwen2-Audio\cite{chu2024qwen2} requires thousands of hours of annotated data as it progresses through multi-task pre-training, instruction-tuning, and direct preference optimization\cite{rafailov2023direct}.


To efficiently create these speech-text pairs, researchers often leverage powerful text-based LLMs with textual metadata from speech\cite{chu2023qwen,chu2024qwen2,hu2024wavllm,lu24c_interspeech}. For instance, LTU\cite{gong2024listen} prompts LLMs to generate open-ended question-answer pairs that cover a diverse array of tasks, which has become a common approach in SLM development. On the other hand, DeSTA\cite{lu24c_interspeech} proposes creating a speech caption dataset from this metadata, allowing the SLM to learn from multifaceted speech descriptions before moving to the instruction-tuning stage. However, these automatic pipelines often incorporate specific criteria in their prompts to guide the generation process, such as in-context examples or length control, and then post-process results into paired data. We believe these commonly used pipelines can potentially introduce text mismatches against the original LLM. In contrast, our approach focuses on minimizing textual mismatches by allowing the LLM to generate responses freely based on the input context.

In the context of preserving the original capabilities of LLMs while adapting them to the speech domain, studies like AudioChatLlama \cite{fathullah2024audiochatllama} and BLSP \cite{wang2023blsp,wang2024blsp} explore the continuation writing behavior of LLMs by feeding text transcriptions from speech and collecting the responses as training targets. However, this approach often overlooks the multifaceted nature of speech, limiting its focus to content-related tasks such as automatic speech recognition or speech translation. In contrast, our approach aims to cover a wide range of speech metadata for a broader understanding of speech information.




{\begin{table*}[t]
\caption{Result on Dynamic-SUPERB and Air-Bench-Chat.}
\begin{center}
\begin{tabular}{lccccccc}
\toprule
\multirow{2}{*}{\textbf{Models}} &\multicolumn{6}{c}{\textbf{Dynamic-SUPERB}} & \textbf{AIR-Bench-Chat} \\

& \textbf{CON} & \textbf{SEM} & \textbf{PAR} & \textbf{DEG} & \textbf{SPK} & \textbf{ALL} & \textbf{Speech} \\
\midrule
\textit{Cascade baselines} \\
\quad ASR + Llama3 (Ours) & 71.45 & 51.52 & 15.07 & 36.00 & 41.70 & 43.59 & 7.01 \\
\quad Specialized Models + Llama3 (Ours) & 82.32 & 63.08 & 25.71 & 59.61 & 40.50 & 58.31 & 7.32 \\

\midrule
\textit{End-to-end systems} \\
\quad LTU-AS\cite{gong_ltuas} & 43.95 & 36.00 & 17.14 & 37.53 & 40.20 & 36.11 & - \\
\quad Salmonn\cite {tang2023salmonn} & 52.00 & 50.75 & 24.50 & 28.16 & 33.20 & 36.44 & 6.16 \\
\quad BLSP-emo\cite{wang2024blsp} & 66.09 & 53.92 & 11.50 & 27.03 & 30.30 & 37.42 & - \\
\quad WavLLM\cite{hu2024wavllm}  & 53.31 & 51.00 & 24.60 & 36.83 & 22.24 & 39.07 & - \\
\quad Qwen-Audio \cite{chu2023qwen} & 61.77 & 47.17 & 28.64 & 30.95 & 41.40 & 40.79 & 6.47 \\
\quad BLSP \cite{wang2023blsp} & 51.82 & 58.25 & 36.00 & 42.76 & 44.80 & 46.00 & 6.17 \\
\quad Qwen2-Audio \cite{chu2024qwen2} & 77.64 & 59.17 & 29.21 & 43.58 & \textbf{47.90} & 51.69 & \textbf{7.18} \\

\quad DeSTA2 & \textbf{79.41} & \textbf{59.42} & \textbf{43.14} & \textbf{51.63} & 42.50 & \textbf{56.78} & 7.16 \\

\bottomrule
\
\end{tabular}
\vspace{-3em}
\label{table:main_results}
\end{center}
\end{table*}}

\section{Method}


\subsection{Dataset construction}
In the dataset construction stage, we aim to create speech-text pairs with minimal textual discrepancies between the underlying LLM used in the SLM and the constructed training data. While LLMs cannot process speech directly, we have observed their proficiency in interpreting textual metadata. Our approach begins by collecting a diverse array of meta-information from speech. We employ annotations from existing datasets and utilize several specialized speech models to enrich the comprehensiveness of metadata. This process encompasses various aspects of speech, including speaking characteristics, speaker information, spoken content, and audio quality metrics.

As depicted in Figure \ref{fig:model}(Left), we encapsulate all relevant information from the speech into a structured format termed a "seed transcript". This format integrates text transcriptions with paralinguistic attributes, following a pattern such as: \textit{"[00:00:00-00:00:03] How are you? (Gender: Female, Emotion: Happy...)"}. Subsequently, we input the seed transcript into the LLM alongside the prompt \textit{"What can you hear from the audio?"} to generate comprehensive speech descriptions. These generated responses to the input context then serve as our training target for the end-to-end SLM model.

\subsection{Model architecture and training strategy}

Our model architecture, illustrated in Figure.\ref{fig:model}(Right), integrates a pre-trained Whisper model\cite{radford2023robust} with an instruction-tuned Llama3\cite{dubey2024llama}. To preserve the inherent capabilities of both pre-trained models, we freeze their parameters during training. A modality adapter with randomly initialized weights bridges the gap between the two models, mapping the output representation from the Whisper encoder to the Llama3 input space.
Specifically, we employ a Qformer\cite{pmlr-v202-li23q,lu24c_interspeech} to extract speech features from the intermediate layers of the Whisper encoder. These features are then aggregated using a weighted sum operation with learnable weights. Finally, a linear projection layer maps the resulting features to match Llama3's input dimensions. During training, continuous speech features from the modality adapter and text transcription from Whisper decoder are concatenated and directly substitute the seed transcript in the template to maintain the same input context. The model undergoes end-to-end optimization using next-token prediction loss.

To sum up, the dataset and training strategy have three key advantages: first, since the dataset is constructed by the same LLM, the resulting SLM doesn't waste any trainable parameters in learning specific response formats or addressing annotation biases. Second, with the descriptive prompt, the model tends to generate comprehensive descriptions of speech, rather than hallucinating meaningless responses based solely on the spoken content\cite{fathullah2024audiochatllama}. Finally, this general idea can be extended to other prompts, while we keep the setup simple and use only one prompt to demonstrate the efficiency and effectiveness of the proposed approach.






\section{Experiment}
{\begin{table}[htbp]
\caption{Statistics of combined dataset.}
\begin{center}
\begin{tabular}{lccc}
\toprule
\textbf{Dataset} & \# Audios & \# Captions & Duration(hours) \\

\midrule

AccentDB & 16874 & 16874 & 19.27 \\
Dailytalk & 20000 & 20000 & 18.17\\
IEMOCAP & 4150 & 20000 & 5.17 \\
PromptTTS & 20000 & 20000 & 38.54\\
VCTK & 20000 & 20000 & 19.90\\
VoxCeleb & 20000 & 20000 & 45.83 \\
Mixed noise\&reverb & 7214 & 7214 & 8.04 \\

\midrule

All & 108238 & 124088 & 154.95\\

\bottomrule
\
\end{tabular}
\vspace{-3em}
\label{table:dataset}
\end{center}
\end{table}}

\subsection{Training dataset}

We leverage datasets such as AccentDB \cite{ahamad-anand-bhargava:2020:LREC}, DailyTalk \cite{lee2023dailytalk}, IEMOCAP \cite{busso2008iemocap}, PromptTTS \cite{guo2023prompttts}, VCTK \cite{yamagishi2019vctk}, and VoxCeleb \cite{nagrani17_interspeech}, each offering expressive speech annotations. We also create a dataset augmented with noise and reverberation to enhance diversity. 

In addition, we use specialized models including gender classification\footnote{https://huggingface.co/alefiury/wav2vec2-large-xlsr-53-gender-recognition-librispeech}, emotion recognition\cite{ma2023emotion2vec}\footnote{https://huggingface.co/emotion2vec/emotion2vec\_plus\_large}, automatic speech recognition\cite{radford2023robust}\footnote{https://huggingface.co/openai/whisper-large-v3}, signal-to-noise ratio (SNR) prediction\cite{lavechin2023brouhaha}\footnote{https://huggingface.co/pyannote/brouhaha}, and C50 prediction\footnotemark[5] models to capture information beyond the original annotations. The resulting dataset encompasses 12 attributes: gender, age, accent, emotion, pitch, volume, speaking speed, SNR level, C50 value, duration, intent, and spoken text. For the generation process, we utilize the Llama3-8B-Instruct and employ sampling decoding with both temperature and top\_p values at 1. To maintain dataset balance, we randomly select training samples from each dataset. Table \ref{table:dataset} presents the statistical overview of our combined dataset.


\subsection{Model specification}

We utilize Llama3-8B-Instruct and Whisper-small (244M parameters) as our foundational components. The modality adapter consists of a 2-layer transformer decoder block that employs 64 trainable vectors as queries to extract speech features from the intermediate hidden state of encoder. As depicted in Figure \ref{fig:model}(Right), we fix Whisper and Llama3, updating only the modality adapter during training. The trainable parameters is 22.3M. Text transcriptions and instructions are transformed using embedding layer of Llama3 and subsequently concatenated with the extracted speech features. We implement our model using the NeMo toolkit \cite{kuchaiev2019nemo} and Hugging Face Transformers library \cite{wolf2020transformers}. The models are trained for 10 epochs using the Adam optimizer with a cosine annealing scheduler and 2000 warm-up steps. We utilize 2 NVIDIA A40 GPUs for training, with a global batch size of 16 and a learning rate of 1e-4.


\subsection{Evaluation setup}

We employ Dynamic-SUPERB \cite{huang2023dynamic} and AIR-Bench-Chat \cite{yang2024air} to evaluate the model performance. For both benchmarks, the model needs to comprehend instructions to perform specific tasks. Dynamic-SUPERB is a crowdsourced benchmark comprising 48 speech-related classification tasks with human-created instructions and answers. These tasks can be categorized into 5 dimensions: content (CON), semantic (SEM), paralinguistic (PAR), degradation (DEG), and speaker (SPK). Generally, tasks in the content and semantic dimensions can be solved with linguistic information, while degradation and speaker dimensions require more paralinguistic information. It uses GPT-4o\footnote{gpt-4o-2024-05-13} as the judge to evaluate the accuracy of each task. AIR-Bench-Chat, on the other hand, is an LLM-generated benchmark derived from speech metadata that focuses on open-ended questions. It utilizes GPT-4\footnote{gpt-4-0125-preview} to assess the agreement (on a scale of 1 to 10) between the ground truth label and the model-generated output.



\section{Results}

\subsection{Results on Dynamic-SUPERB and AIR-Bench-Chat}

Table \ref{table:main_results} presents a comprehensive comparison of our proposed model against various baselines and  SLMs on the Dynamic-SUPERB and AIR-Bench-Chat.
At first glance, DeSTA2 exhibits remarkable performance across diverse speech tasks. Notably, these impressive outcomes are achieved \textbf{without} any task-specific finetuning. In contrast to previous SLMs, such as Qwen2-Audio, which require multi-stage training with extensive annotated pair data, our approach is more data-efficient and demands less human effort.
On Dynamic-SUPERB, we achieve an overall accuracy of 56.78\%, significantly outperforming other end-to-end systems. For AIR-Bench-Chat, DeSTA2 attains a competitive score of 7.16, closely trailing the current state-of-the-art Qwen2-Audio (7.18).
Furthermore, comparing across task categories in Dynamic-SUPERB, our model consistently outperforms other end-to-end models in most dimensions, except for the speaker dimension. In particular, the speaker verification task requires the model to compare two audio samples, which is an aspect not covered in this study.
Compared to cascade baselines, our model surpasses ASR+Llama3 and shows comparable performance to the strong specialized baselines. This suggests the effectiveness of our approach in aligning end-to-end SLM without compromising the original LLM's ability.







\subsection{Ablation study on data construction methods}

To compare the effectiveness of our method against various data construction approaches based on the same metadata, we used the system prompt from \cite{lu24c_interspeech} with Llama3-8B-Instruct to generate speech captions, denoted as DeSTA. Additionally, we employed a more powerful Llama3-70B-Instruct to create both three and one open-ended question-answer pairs for each speech sample \cite{gong2024listen}, denoted as Open QA (3) and Open QA (1), respectively. These baseline approaches potentially introduce textual mismatches during the data construction process.

As demonstrated in Table \ref{table:ablation_results}, the results indicate that training with DeSTA or seed transcripts as targets can be problematic due to significant mismatches in the training data. The resulting model tends to ignore instructions and produce only captions or transcripts during inference \cite{lu24c_interspeech}, which is considered a redundant response in the evaluation.
Secondly, training with open-ended QA pairs shows more reasonable performance. However, while using three QA pairs covering more aspects of speech requires three times as many updates compared to using a single QA pair for each audio, the results reveal an interesting trade-off. With more updates, the model improves in the degradation category but declines in content and semantic categories, indicating a trade-off between language ability and speech understanding.
In contrast, DeSTA2 alleviates the aforementioned issues and demonstrates significantly better performance than other construction methods. It is worth highlighting that our approach employs only a single task (speech captioning) with a unified prompt to generate all speech-text pairs. DeSTA2 challenges the prevailing belief that finetuning SLMs requires diverse instruction data encompassing a wide range of tasks. We provide insights into SLM development by demonstrating that a proper data construction process with rich speech metadata can already build a robust and versatile SLM.

{\begin{table}[]
\caption{Ablation results on varios dataset construction methods.}
\begin{center}
\begin{tabular}{lccccc|c}
\toprule
\multirow{2}{*}{\textbf{Models}} &\multicolumn{6}{c}{\textbf{Dynamic-SUPERB}} \\

& \textbf{CON} & \textbf{SEM} & \textbf{PAR} & \textbf{DEG} & \textbf{SPK} & \textbf{ALL} \\
\midrule
DeSTA & 4.23 & 9.25 &14.86 &7.34 &0.40 &7.24 \\
Seed transcript & 63.50 & 49.92 &27.50 & 18.03 & 11.40 & 33.13 \\
Open QA (3) & 64.95 & 51.75 & 22.86 & \textbf{54.03} & 40.90 & 50.33 \\
Open QA (1) & 77.27 & 53.83 & 23.07 & 46.50 & 41.80 & 50.56 \\
DeSTA2 & \textbf{79.41} & \textbf{59.42} & \textbf{43.14} & 51.63 & \textbf{42.50} & \textbf{56.78} \\

\bottomrule

\end{tabular}
\vspace{-1.5em}
\label{table:ablation_results}
\end{center}
\end{table}}

\subsection{Results on finetuned with speech instruction-tuning dataset}

{\begin{table}[]
\caption{Comparison between models finetuned with in-domain instruction-tuning dataset.}
\begin{center}
\vspace{-1em}
\begin{tabular}{lccccc|c}
\toprule
\multirow{2}{*}{\textbf{Models}} &\multicolumn{6}{c}{\textbf{Dynamic-SUPERB}} \\

& \textbf{CON} & \textbf{SEM} & \textbf{PAR} & \textbf{DEG} & \textbf{SPK} & \textbf{ALL} \\
\midrule

Whisper-LLM\cite{huang2023dynamic} & 64.41 & 33.17 & 41.79 & 69.79 & \textbf{79.00} &60.85 \\
DeSTA\cite{lu24c_interspeech} & 91.27 & \textbf{73.00} &51.21 & 61.87 &54.00 & 67.63 \\
FT-only (Ours) & 92.23& 68.92& 53.14& 67.87&  62.40&  70.86 \\
DeSTA2 & \textbf{96.00}&  71.75 &  \textbf{56.86}& \textbf{70.26}& 70.80 & \textbf{74.45}  

\\

\bottomrule
\
\end{tabular}
\vspace{-3em}
\label{table:finetune_results}
\end{center}
\end{table}}
As an end-to-end system, our model can be further enhanced through finetuning with task-specific datasets. Table \ref{table:finetune_results} presents the results of finetuning our model with in-domain Dynamic-SUPERB training set and incorporating LoRA\cite{hu2021lora} (Rank=8), where FT-only denotes we finetune the modality adapter from scratch without speech caption training. This additional finetuning phase enables the model to acquire task-specific features and patterns that may not have been fully captured during initial training. These results highlight the model’s adaptability and its capacity to serve as a foundation model for specialized speech processing applications.



\subsection{Examples on complex reasoning ability}

{\begin{table}[]
\caption{Results on paralinguistic complex reasoning.}
\begin{center}
\vspace{-1em}
\begin{tabular}{l}
\toprule

\textbf{Input Audio:}
\\Text: Thank you for calling. Goodbye, Duration: 2.8s, Emotion: Happy


\\ \midrule
\textcolor{gray}{\textbf{Following output format}}\\
\textcolor{blue}{\textbf{Q:} What is the emotion of the speaker? Answer in all capital letters. }\\
\textbf{DeSTA2:} HAPPINESS\\
\textbf{Qwen2-Audio:} THANK YOU FOR CALLING GOODBYE.\\

\midrule
\textcolor{gray}{\textbf{Chain-of-thought reasoning}}\\
\textcolor{blue}{\textbf{Q:} Analyze the audio, then answer "What is the emotion of the speaker?"}\\
\textbf{DeSTA2:} Based on the audio, the speaker's emotion is "happy". The\\response 
is a polite and cheerful goodbye, which indicates a positive and \\friendly
tone. The tone is upbeat and warm, suggesting that the speaker \\ 
is in a good mood and is satisfied with the outcome of the conversation. \\
\textbf{Qwen2-Audio:} Happy \\

\midrule
\textbf{\textcolor{gray}{
Math with chain-of-thought reasoning
}}\\
\textcolor{blue}{\textbf{Q:} Calculate the words per second rate, think step by step.}\\
\textbf{DeSTA2:} Let's break it down step by step: \\1. The speech lasts for 2 seconds: [00:00:00 - 00:00:02]\\2. The number of words is:"Thank you for calling. Goodbye."(5 words)\\3. To find the words per second rate, we need to divide the number of \\words by the time it took to say the words:\\Words per second = Total number of words / Time taken (in seconds)\\= 5 words / 2 seconds = 2.5 words per second\\Therefore, the words per second rate is 2.5 words\\
\textbf{Qwen2-Audio:} First, divide the total number of words by the number of\\seconds to get the average words per second rate. In this case,there are\\6 words in the speech and it took 2 seconds to say them. So the average\\words per second rate is 3 words per second. \\

\bottomrule
\end{tabular}
\vspace{-2em}
\label{table:real_examples}
\end{center}
\end{table}}


Since current LLM are highly capable in following instruction or perform chain-of-thought reasoning\cite{wei2022chain}. As illustrates in Table \ref{table:real_examples}, we found our model not only accurately identifies paralinguistic information in speech but also adeptly follows specific instructions and engages in sophisticated reasoning tasks. In contrast, responses from Qwen2-Audio do not always follow the instructions. This flexibility is a direct inheritance from the text-based LLM, extending well beyond the scope of the speech-text training data.
Our training approach successfully preserves the advanced capabilities from underly LLM and seamlessly applies them to the speech domain. Notably, this ability is rarely observed in previous SLMs, underscoring a significant advantage of our method.

\section{Conclusion}
In this work, we show that it is possible to develop an instruction-following SLM without speech instruction-tuning data. We present an automatic and scalable approach that uses speech captions generated by an underlying LLM to train the end-to-end SLM. This approach greatly simplifies the data construction process and alleviates potential forgetting problems. Our model extends the advanced capabilities from LLM to the speech domain and we conduct extensive comparisons against various SLMs and data construction methods on Dynamic-SUPERB and AIR-Bench-Chat. This success provides a more data-efficient path toward universal speech language models.



\section{Acknowledgment}
The authors thank NVIDIA Taiwan AI R\&D Center for the TRDC budget support and contributions to this research.

\bibliography{mybib}

\end{document}